\documentstyle[11pt,aaspp4]{article}

\def\simg{\mathrel{\hbox{\rlap{\lower.55ex \hbox {$\sim$}}
                   \kern-.3em \raise.4ex \hbox{$>$}}}}
\def\siml{\mathrel{\hbox{\rlap{\lower.55ex \hbox {$\sim$}}
                   \kern-.3em \raise.4ex \hbox{$<$}}}}
\def\Mesz{M\'esz\'aros~}
\def\Pacz{Paczy\'nski~}

\def\beq{\begin{equation}}
\def\enq{\end{equation}}
\def\bea{\begin{eqnarray}}
\def\ena{\end{eqnarray}}
\def\bec{\begin{center}}
\def\enc{\end{center}}
\def\etal{{\it et al.}}

\def\ergsi{\hbox{erg s$^{-1}$}}
\def\cmcui{{\rm cm}$^{-3}$}
\def\msun{M_\odot}
\def\Q50{L_{50}}
\def\L50{L_{50}}

\def\refe{\reference}
\def\cm{{\rm cm}}
\def\cmsqi{{\rm cm}^{-2}}
\def\cmcui{{\rm cm}^{-3}}
\def\gcmcui{{\rm g~cm}^{-3}}
\def\si{{\rm s}^{-1}}
\def\para{\parallel}
\def\G0{\Gamma_0}
\def\rhoh8{\rho_{H,-8}}
\def\th0{\theta_0}
\def\r06{r_{0,6}}
\def\rfe9{r_{Fe,9}}
\def\rhe11{r_{He,11}}
\def\rh13{r_{H,13}}
\def\Ki{K_i}
\def\Ko{K_o}
\def\Eb51{E_{b,51}}

\begin{document}
{\footnotesize \hfill ApJL, accept. 6/20/01; astro-ph/0104402}

\title{ Collapsar Jets, Bubbles and Fe Lines}


\author{
P. \Mesz$^{1}$ \&
M.J. Rees$^2$ 
}
\bec
\smallskip\noindent
$^1${Dpt. of Astronomy \& Astrophysics, Pennsylvania State University,
University Park, PA 16803}\\
\smallskip\noindent
$^2${Institute of Astronomy, University of Cambridge, Madingley Road, Cambridge
CB3 0HA, U.K.}\\
\enc

\begin{abstract}
In the collapsar scenario, gamma ray bursts are caused by relativistic  
jets expelled along the rotation axis of a collapsing stellar core.  
We discuss how the structure and time-dependence of such jets depends on 
the stellar envelope and central engine properties, assuming a steady jet injection.  
It takes a few seconds for the jet to bore its way through the stellar core;
most of the energy output during that period goes into a cocoon of relativistic 
plasma surrounding the jet. This material subsequently forms a bubble of  
magnetized plasma that takes several hours to expand, subrelativistically, 
through the envelope of a high-mass supergiant. 
Jet break-through and a conventional burst would be expected not only in 
He stars but possibly also in blue supergiants.
Shock waves and magnetic dissipation in the escaping bubble 
can contribute a non thermal UV/X-ray  afterglow, and also excite Fe line 
emission from thermal gas, in addition to the standard jet deceleration 
power-law afterglow. 

\end{abstract}


\section{Initial Model}
\label{sec:mod}

Collapsars seem a good model for the `long' (Beppo SAX) gamma-ray bursts (GRB), 
though not for the short ones (e.g. \Pacz, 1998; Woosley, 1999). Numerical
simulations of the progenitor collapse have been made, e.g. by MacFadyen, 
Woosley \& Heger (2001) and references therein. We will use as an example 
their progenitor model A25, a pre-supernova star of initial mass $M_i=25\msun$, 
reduced to $14.6\msun$ by mass loss, with restricted semi-convection and 
rotation. This has a red supergiant structure, with a $\sim 2\msun$ Fe core 
of radius $r_{Fe}\sim 7\times 10^8$ cm, an $8.4\msun$ He core extending out 
to $r_{He}\sim 10^{11}$ cm, and a hydrogen envelope out to $r_H\sim 10^{13}$ cm.  
The average specific angular momentum within the He core is $j\sim 5\times 
10^{16}\cmsqi\si$, with values $\sim 50\%$ higher in the equator.  However, 
the results we shall discuss also apply qualitatively to other cases, e.g.
models with larger initial masses, which may have lost more or all of their 
H envelope and whose outer radii are $\sim r_{He} \sim 10^{11}$ cm.

A central black hole (BH) of several $\msun$ forms after collapse of the 
Fe core, in a free-fall $t_{ff,Fe}\sim 1$ s, and the subsequent jet evolution 
was calculated numerically, non-relativistically by MacFadyen \etal 2001, and 
relativistically by Aloy \etal, 2000. These 
calculations assume that thermal energy generated by the accreting BH is 
deposited in a conical region between 200-600 Km out along the rotation axis.
In Aloy \etal's calculation, using a star similar to A25 but without an H 
envelope, the jet bulk Lorentz factor reaches intermittently values 
$\bar {\Gamma_j}\siml 5-30$ in rarefied portions of the flow not near the 
head of the jet, while throughout most of the star the average jet advance 
speed is $\siml 10^{10} \cm\si$. This would have interesting implications 
for the ability of the collapsar to make a GRB: jets fed by a central engine 
whose duration (say, $t_j \siml 10-100$ s) is shorter than a crossing time 
would not break through, and would not produce a GRB (MacFadyen \etal, 2001).

Some aspects of the dynamics  may, however, modify the development of the
jet in a rapidly rotating collapse, particularly if there is a delay
between the   formation of a collapsed core and the initiation of a jet.
The effect of rotation is to create a funnel  of a determined shape along
the rotation axis, within which the density  becomes substantially lower
than the equatorial. This is because,  within this
funnel, material drains into the hole on a free-fall timescale, rather than
being centrifugally supported. Jets that propagate out after this funnel
has developed would, at least at small  $r$,  develop differently from those
studied by  Aloy \etal (2000) and MacFadyen \etal, 2001.

The pre-collapse central density of the pre-SN model A25 is  
$\rho\simeq 10^{12}\cmcui$, and out to the edge of the He core 
$r_{He}\sim 1.5\times 10^{11}$ cm it scales roughly as 
$\rho\propto r^{-3}$, while the (radiation dominated) pressure 
is $p\propto \rho^{4/3}\propto r^{-4}$.  Loss of pressure support leads 
in a few seconds (the free-fall time from the radius enclosing the 
innermost few solar masses) to a BH, girded by a centrifugally supported 
torus, which heats up to virial temperatures and cools through neutrino 
losses. Mass infall proceeds freely along the rotation axis, 
leading via the continuity equation to $\rho \propto r^{-3/2}$.
Thus, a few seconds after collapse, the gas density {\it inside} the
funnel near $r_0 \simg 10^6\cm$ may be approximated by the pre-collapse
density at the edge of the Fe core at $r_{Fe}\sim 7\times 10^8$ cm
multiplied by the free fall compression ratio,
$
\rho_{i,0}\sim \rho_{Fe} (r_{Fe}/r_0)^{3/2} \sim 5\times 10^{10}  
                        \Ki \rfe9^{3/2} \r06^{-3/2} ~\gcmcui~.
$
The gas {\it outside} the funnel, which takes a free-fall
time to form, initially was in approximately radial free-fall, until it
starts to pile up near the equatorial plane.
A jet, fed near $r_0$ by  a strong Poynting flux 
from the inner disc or the hole itself, will then be channeled along the 
rotation axis.  In reality, the jet
power may fluctuate; so also may its baryon content, due to entrainment, or
to unsteadiness in the acceleration process at the base of the jet.
In the following section, however, we treat the jet as being steady.

\section{Jet Dynamics}
\label{sec:jet}

The dynamics of the collapsing rotating stellar gas with a BH at the bottom
implies that there is a critical specific angular momentum $j_0 \sim
10^{16}\cmsqi\si$.
Since the transverse velocity at radius $r$ cannot exceed the escape
velocity
($\propto r^{-1/2}$) this implies (Fishbone \& Moncrief, 1976) that, after
the BH forms, there is a funnel around the rotation axis, within which
there can be no axisymmetric equilibrium. For $r\gg r_0$ the
funnel opening angle has a paraboloid shape,
\beq
\theta_\ast \sim \theta_0(r/r_0)^{-1/2}~.
\label{eq:thetaast}
\enq
Matter falling in from the outer edge of the Fe core at $r\sim 10^9\rfe9$
cm will, in a few seconds, result in a considerably lower density within 
the funnel near the BH horizon, creating the conditions for the launching 
of a jet. As inferred GRB and afterglow observational fits, the 
jet must be injected with a dimensionless entropy per baryon $\eta=
(L_j/{\dot M} c^2)\sim 10^2\eta_2$, where $L_j$ is kinetic luminosity,
$\dot M$ is mass loss rate, $c=3\times 10^{10}\cm\si$. This implies a low
density, highly relativistic jet, injected by the central engine  at the 
base of the outflow.  For the purposes of calculation we take an initial 
Lorentz factor $\G0\sim 1$ and an initial jet opening angle $\th0 \sim 1$ 
(in radians; e.g. Aloy, \etal, 2000). We assume a
steady, adiabatic jet (internal shocks would occur, if at all, at radii
$\simg 10^{12}-10^{13}$ cm, while small scale, Thomson-thick dissipation
results in trapped radiation which is effectively adiabatic).
From the relativistic Bernoulli equation (e.g. Begelman \etal, 1984),
the jet, of opening angle $\theta_j$, gradually converts at $r> r_0$ its
internal energy into bulk kinetic energy with $\Gamma_j >1$,
\beq
\Gamma_j \propto (r\theta_j)~= \cases{
  \G0 (r/r_0)^{1/2}  & ($\theta_j \propto r^{-1/2},~~~r\siml r_f$); \cr
  \G0 (r_f/r_0)^{1/2}(r/r_f) & ($\theta_j=\hbox{const.},~~r\simg r_f$),
  \cr}
\label{eq:Gammaj}
\enq
where $r_f$ is defined in equation (\ref{eq:rf}).  Thus the jet
internal Lorentz factor increases as $\Gamma_j \propto r^{1/2}$, until the
jet opening angle behavior changes or until a maximum saturation
value $\Gamma_j \sim \eta=10^2\eta_2 =$ constant. The
funnel behavior $\theta_j \propto r^{-1/2}$ can 
be maintained only so long as transverse balance holds between the
jet pressure (assumed isotropic) and the stellar pressure outside the
funnel. The radius where this occurs depends on specific assumptions about jet
production. 
(Note that ours differ from Aloy~ \etal, 2000; we assume that 
jet propagation starts after a {\it parabolic} funnel has  developed; 
and we postulate a jet injection with high internal energy to rest mass 
ratio, constant throughout the central energy life).

A key aspect is that the $\Gamma_j \gg 1$ jet slows 
down abruptly in a narrow layer near the head of the jet, which advances 
into the star with an initially sub-relativistic velocity.  The jet head 
velocity $V_h < c$ is given by the longitudinal balance between the 
jet thrust per unit area $p_{j\para}\sim (L_j/2\pi\theta^2 c  r^2)$ 
and the ram pressure of the funnel material ahead of it, 
$p_{e} \sim \rho_i(r) V_h^2$.  This gives at $r_0$ a fiducial jet 
head velocity 
\beq
V_{h0}=V_h(r_0) \sim 10^8 \L50^{1/2}\Ki^{-1/2} \th0^{-1} \r06^{-1/4} 
\rfe9^{-3/4}~\cm~.
\label{eq:Vh0}
\enq
(This is a crude approximation of the dynamics near $r_0$,
because of the possible rapid infall of the gas, and the complexities 
of jet formation; however, it sets the scale for the values of $V_h$ 
at $r\gg r_0$ where the approximations are likely to be better).

In front of the contact discontinuity between the jet and the stellar gas 
there is a thin layer of shocked stellar gas moving ahead with velocity 
$\sim V_h$ into the star. Behind the contact discontinuity there is a shock 
where the relativistic jet material, with  $\Gamma_j\gg 1$,   is slowed to 
a velocity of order $V_h$. Between this inner shock and the contact 
discontinuity is a layer of shocked relativistic plasma which also moves 
with velocity $\sim V_h$. Both the unshocked $\Gamma_j\gg 1$ jet gas and 
the (higher pressure) shocked relativistic gas   remain, at low radii, 
trapped in the transverse direction by the paraboloidal funnel $\theta_j
\propto r^{-1/2}$, sideways expansion being prevented by the much higher 
external pressure (and inertial confinement, beyond $r_f$ given by 
equation [\ref{eq:rf}]).  The  shocked jet gas would have such a high 
electron density (augmented by pairs) that radiation would be trapped, 
so that it behaves adiabatically even for short radiative cooling times.
The shocked material occupies a volume  $(c/V_h)$ larger than the jet 
material itself, so the relativistic shock would be located $(c/V_h)^{1/2}$ 
closer in than the head of the jet (since the volume within a thin 
paraboloid goes as $r^2$).
In the stellar frame, the pressure of the unshocked relativistic jet
in the transverse direction is, from the relativistic Bernoulli equation,
$p_{j\perp}\sim (1/3)L_j/(2\pi r^2 \theta^2 c \Gamma_j^2) \sim 4\times 
10^{25}\L50 \r06^{-2}\th0^{-2}\G0^{-2} (r/r_0)^{-2}$ cgs, as long as 
$\theta_j\propto r^{-1/2}$.  At $r\siml r_{Fe} \sim 10^9\rfe9$ cm 
a few seconds after the collapse, we can approximate the stellar gas 
density profile outside the funnel as given by free-fall, $\rho \propto 
r^{-3/2}$, with a (radiation dominated) pressure $p\propto \rho^{4/3} 
\propto r^{-2}$. Using MacFadyen et al's 2000 pressure at $r_{Fe}$, 
the value near $r_0$ is
$
p_{o,0}\sim p_{Fe}(r_{Fe}/r_0)^2 \sim
  4\times 10^{29} \Ko^{4/3} \rfe9^2 \r06^{-2} ~{\rm dyne}~{\rm cm}^{-2}~,
$
where $K_o\sim 1$, and 
$p_{e} = p_{e,0} (r/r_0)^{-2}\sim 4\times 10^{29}\Ko^{4/3}\rfe9^2$ cgs.
The external pressure at low radii is much larger than that of the 
unshocked $\Gamma_j \gg 1$ jet. The shocked jet gas pressure is larger 
than that of the unshocked jet gas, since it moves with $V_h < c$ 
and is not affected by the $\Gamma_j^{-2}$ factor. The value of
$p_{h\perp} \sim (1/3) L_j/(2\pi r^2 \theta^2 c) \sim p_{h\para}$ is 
comparable to the external ram pressure $\rho_e V_h^2$, and decays as 
$p_{h\perp}\sim 5\times 10^{26} \L50\th0^{-2} \r06^{-2} (r / r_0)^{-1}$ cgs, 
whereas the (initially higher) external pressure is $\propto r^{-2}$.  
They become equal at a radius 
$r_f/r_0\sim 10^3\L50^{-1}\Ko^{4/3}\th0^2\rfe9^2$ or
\beq
r_f\sim 10^9 \L50^{-1}\Ko^{4/3}\th0^2\rfe9^2 \r06~{\rm cm}~,
\label{eq:rf}
\enq
which is close to the radius $r_{Fe}$.
(The exact details of the external pressure profile below $r_{Fe}$ 
are not crucial, as long as it depends on $r$ more steeply than the 
shocked-jet gas pressure $1/r$). 
At the radius $r_f$, the funnel opening angle is then
\beq
\theta_f =(r_f/r_0)^{-1/2}\sim 3\times 10^{-2} 
\L50^{1/2}\Ko^{-2/3}\th0^{-1}\rfe9~.
\label{eq:thetaf}
\enq

For $r> r_f$ the transverse pressure of the shocked jet 
material  exceeds the external stellar pressure, and it can no 
longer be confined inside the paraboloidal funnel $\theta_\ast\propto
r^{-1/2}$.  It spills out  sideways and trails behind the jet, creating  a 
sheath of low density, relativistic material resembling the cocoons 
(cf Scheuer 1974)  that surround the relativistic jets of radio sources. 
In Aloy \etal 2000, based on different initial assumptions, a cocoon appears 
at smaller radii. Cocoons are generally over-pressured, whenever their
boundary moves faster than the sound speed in the surrounding stellar gas.
The cocoon expands in the transverse direction until it reaches 
pressure equilibrium with the external stellar gas.  The jet head advances 
subrelativistically with $V_h <c$. But the  shocked jet material  has an 
internal sound speed $\sim c/3$, so the pressure in the cocoon is equalized
almost instantaneously throughout its entire volume. The unshocked jet gas 
moving with $\Gamma_j \gg 1$ is now constrained in the transverse direction 
by the pressure of the cocoon bubble, which in turn is in equilibrium with 
the external stellar pressure, $p_b \sim p_e \propto r^{-n}$ 
(where for $r\simg r_{Fe}$ we take the approximate pre-collapse pressure 
profile $n \sim 4$, e.g. \S 1). The relativistic jet opening angle $\theta_j$ 
adjusts itself to satisfy $p_{j\perp}\propto L_j/(r^2 \theta_j^2 \Gamma_j^2)
\propto p_b\propto p_e \propto r^{-n}$, and $\Gamma_j \propto (r\theta_j)$, 
so
\beq
\theta_j \propto r^{(n-4)/4} \propto \cases{
                                 r^{-1/2}  & for $r\siml r_{Fe}$; \cr
                         {\rm constant}    & for $r\simg r_{Fe}$. 
\cr}
\label{eq:thetaj}
\enq
Thus, above $r_f\sim r_{Fe}$, the jet becomes ballistic, with $\theta_j
\sim \theta_f \sim$ constant, which from equation (\ref{eq:thetaf}) is a 
few degrees.  From $p_{j\para} \propto L_j/(r^2\theta_j^2) \propto 
\rho_e V_h^2$, the jet head advance velocity is 
\beq
V_h \propto r^{n/8} \propto \cases{
                             r^{1/4}  & for $r\siml r_{Fe}$; \cr
                             r^{1/2}  & for $r\simg r_{Fe}$. \cr}
\label{eq:Vh}
\enq
The head reaches $r_f\sim r_{Fe}$ in approximately a second, with 
$V_h(r_f) \sim 10^9 \L50^{1/2} \Ki^{-1/2} \th0^{-1} \r06^{-1/2} \rfe9^{-1/2}$ cm/s. 
The bubble surrounding the jet would be cigar-shaped (cf Scheuer 1974) 
in a uniform pressure external medium. However, a stellar pressure $p_e \propto 
r^{-n}$ will result in a bottom-up pear shape (also in Aloy \etal 2000). 
Balancing the bubble pressure $p_b \sim L_j t/(\pi r^2 \theta_b^2 V_h t)$ 
and the external pressure $p_e\propto r^{-n}$, the bubble opening angle 
$\theta_b$ is 
\beq
\theta_b \propto r^{(n-2)/2}V_h^{-1/2}\propto r^{(7n-16)/16}\propto \cases{
                                     r^{-1/8}  & ($r\siml r_{Fe})$;\cr
                                     r^{3/4}   & ($r\simg r_{Fe})$.\cr}
\label{eq:thetab}
\enq
For the model used here the second line is applicable, since the 
external pressure equals the jet pressure at $r_f\sim r_{Fe}$, and it is 
only beyond this equality radius that the  shocked relativistic plasma can  
expand sideways to form a cocoon.

At $r > r_f \sim r_{Fe}\sim 10^9\rfe9$ cm, for the timescales of a few 
seconds of interest here, the pre-collapse profile $p_{e} \propto r^{-n}$ 
applies with $n\sim 4$. From equation (\ref{eq:Vh}) the jet head velocity 
$V_h\propto r^{1/2} \to c$ by the time it reaches $r_{He}\sim 
10^{11}\rhe11$ cm, approximately 10 seconds after jet is launched.
For $r> r_f$ the jet opening angle remains constant, 
$\theta_j\sim \theta_f\sim 0.03$. From equation (\ref{eq:Gammaj}), the 
internal Lorentz factor $\Gamma_j \sim \G0(r_f/r_0)^{1/2}(r/r_f) \propto r$, 
and, in the absence of strong internal dissipation,  it reaches its maximum 
saturation value $\Gamma_j \sim 10^2\eta_2$ (where $\eta={\dot L_j}/{\dot M} 
c^2$) at a radius $r_\eta$ where $r_{Fe}<r_\eta <r_{He}$.

If the external pressure changes gradually, causal contact can be maintained 
across the jet with the exterior (ensuring that the jet is able to bend 
smoothly along the funnel) when
$\theta_j~\Gamma_j \siml 1$ for $\Gamma_j < \eta$, and
$\theta_j~\Gamma_j \siml (v_{th}/c)^2$ for $\Gamma_j=\eta $.
Here $v_{th}$ is sound speed, and $(v_{th}/c)^2 \propto r^{-2/3}$ accounts 
for adiabatic cooling after the Lorentz factor saturates. The first case
is approximately satisfied for $r<r_f$, since 
$\theta\Gamma_j \sim \theta_0\Gamma_0\sim 1 \sim $ constant.

The energy concentration in the bubble is not of course as extreme as in 
the inner jet itself, but it is still exceedingly high: even at $r \sim 
10^{11}$ cm, the energy density  would be $\sim 10^{20}$ erg cm$^{-3}$, 
corresponding to  field strengths more than $10^{10}$ G.

At the edge of the He core $r_{He} \sim 10^{11}\rhe11$ cm, which has 
not had time to collapse before the jet reaches it (at $t\sim 10$ s after 
the jet is launched), the density suddenly drops by a large factor, from 
about $\rho_{He}\sim 1 ~\gcmcui$ to $\rho_H \sim 10^{-8} \rhoh8 \gcmcui$.
These values are for the stellar model A25 of MacFadyen \etal, 2001, but 
the behavior is typical of massive stars (Woosley, Langer \& Weaver, 
1993), e.g. supergiants and Wolf-Rayet stars. When there is a 
remaining H envelope beyond the He core, as in model A25, it has a 
drastically lower density, out to an outer radius $r_H \simg 10^{13}\rh13$ cm.  
The sudden, large density drop at the boundary of the He core gives a large 
boost to the jet head Lorentz factor, from $\Gamma_h \sim 1$ to a highly 
relativistic value.  Whether the jet head is relativistic or sub-relativistic
upon arrival at $r_{He}$, beyond this radius
the jet head Lorentz factor is given by balancing the jet thrust to the 
external ram pressure in the (much) reduced density H envelope, using the 
final jet opening angle (\ref{eq:thetaf}). Thus, the head of the jet 
emerges from the He core into the H envelope with a Lorentz factor
\beq
\Gamma_{h,He} \simeq 
~50~\Ko^{1/3}\th0^{1/2}\r06^{1/4}\rfe9^{1/2}\rhe11^{-1/2}\rhoh8^{-1/4}~,
\label{eq:GammahH}
\enq
not far below the limiting value $\eta =10^2\eta_2$. For a slowly
varying H-envelope density $\rho_H\propto r^{-m}$ (where $m\siml 1/2$ 
approximately fits the typical stellar models), the head Lorentz factor 
varies as $\Gamma_h \propto r^{-(2-m)/4}\propto r^{-3/8}$ for $m=1/2$. 
Near the outer edge of the H envelope, its value is $\Gamma_{h,H}\sim 
10~ \Ko^{1/3}\th0^{1/2} \r06^{1/4}\rfe9^{1/2}\rhoh8^{1/4}\rhe11^{-1/8} 
\rh13^{-3/8}$. This is reached in a crossing time $t_H \sim 
r_H/2c\Gamma_{h,H}^2 \sim 2~\Ko^{-2/3}\th0^{-1}\rfe9^{-1}\r06^{-1/2}
\rhoh8^{-1/2} \rhe11^{1/4} \rh13^{7/4}$ s.  
After that, the jet escapes through an exponentially decreasing atmosphere 
into the circumstellar environment, where it would acquire the limiting 
bulk Lorentz factor, $\Gamma_{hf}\sim \Gamma_j \sim \eta =10^2\eta_2$.

\section{Cocoon and Bubble Dynamics and X-ray Lines}
\label{sec:bubble}

After emerging into the low density H envelope at $r > r_{He}\sim 
10^{11}\rhe11$,  the jet head advances ultrarelativistically.
It then no longer produces a cocoon, for two reasons: (i) the 
energy/momentum  ratio of the swept-up material is such that there is no 
`waste energy' (whereas when $V_h$ is subrelativistic only a fraction 
$(V_h/c)$ of the jet energy is used up in pushing the jet head outwards); 
(ii) causality constraints would in any case prevent shocked material from 
expanding sideways through a significant angle.  However,  the relativistic 
material that  accumulated in the cocoon while the jet was advancing inside
$r_{He}$  would, for a jet crossing time $t_{He}\sim 10$ s (\S \ref{sec:jet}), 
amount to $E_b\sim L_j t_{He}\sim 10^{51}\Eb51$ ergs. This is much larger 
than the binding energy of the H envelope.

The ( highly relativistic) jet head crosses the H envelope in a time of 
order $r_H/c$. An observer along the line of sight would see this speeded 
up by the square of the bulk Lorentz factor -- so that, 
unless $M_{env} \simg 5\msun$ or $r_H \simg 10^{13}\cm$, 
a gamma-ray burst could be detected beyond $r_H$ within a few seconds of 
the jet head reaching $r_{He}$.  The cocoon material would itself 
be able to `break out' and accelerate as soon as the jet penetrated into 
the low-density envelope beyond $r_{He}$.   It has a relativistic internal 
sound speed; however, unlike the jet, it does not have a relativistic 
outward motion.   It  therefore, on reaching $r_{He}$, spreads out over a 
wide angle, and  expands through the envelope as a bubble, in approximately 
the same way as an impulsive fireball in a tenuous external medium

The bubble breaks out of the H-envelope
with a velocity $V_{b,H}\sim c (E_b/M_{env} c^2)^{1/2} 
\sim 10^9 (\Eb51/M_{env,\odot})^{1/2}$ cm/s, at a time
$t_{b,H}\sim r_H/V_{b,H} \sim 10^4 \rh13 (M_{env,\odot}/\Eb51)^{1/2}$ s 
after the burst. The black-body temperature of the bubble is $T_b\sim 
10^{6.5} \Eb51^{1/4}\rh13^{-3/4}$ K, so pairs are no longer in equilibrium. 
However, at $r\sim 10^{13}\rh13$ cm the bubble is still Thomson thick, 
since during its build-up as a relativistic spill-over reservoir it acquired 
a baryon load $M_b\sim E_b/\eta c^2 \sim 10^{28} \Eb51 \eta_2^{-1}$ g, hence 
its average particle density is $n_{b,H}\sim 10^{13} ~\Eb51\eta_2^{-1}
\rh13^{-3}~\cmcui$, and its average mass column density is $y\sim M_b/r_H^2
\sim 10^2 ~\Eb51 \eta_2^{-1}\rh13^{-2}$ g cm$^{-2}$.  This is a lower 
limit to the bubble mass: during the production of the cocoon,  
stellar core material could have been mixed in and entrained (Aloy \etal 2000); 
and there would be further mixing with the envelope during the later 
bubble-expansion. Moreover, if the jet which supplied the original 
material was  magnetically-driven, the bubble would still have a 
dynamically-important (and perhaps dominant) magnetic field: its strength 
would then be $B\sim 10^5$ G even after expansion to $10^{13}$ cm.

If the field were tangled, continuing reconnection would lead 
to acceleration of non-thermal electrons. There would also be subrelativistic 
shocks at the bubble boundary,  which compress the envelope gas,
probably inducing  further clumpiness, and a reverse shock that moves into 
the bubble. The minimum electron random Lorentz factor in equipartition
with shocked relativistic protons is $\gamma_m\simg 6\times 10^2$, which in 
an equipartition magnetic field $B\sim 10^5$ G will produce a synchrotron 
UV/X-ray continuum. The lifetime of electrons  in such strong fields is  of 
course very short. However, the dissipation by shocks and reconnection would 
provide a continuous supply of fresh electrons. It is therefore plausible 
that a substantial fraction of the energy stored in the bubble 
(i.e. at least  $10^{51}-10^{52}$ ergs) could be  released in the  few hours 
after the burst.
A magnetic field of $10^5$ G could confine clumps or filaments of gas 
with densities up to  $n\simg 10^{17}\cmcui$, even at keV temperatures. 
Such filaments would be individually optically thick, and could reprocess
a non-thermal UV/X-ray continuum $L_x \sim E_b/t_{b,H}\sim 10^{47}~\ergsi$
arising in the dilute plasma between them.

The UV/X-ray  continuum  would maintain a clump ionization parameter 
$\xi=L_x/n r^2 \sim 10^3-10^4$, and the $Fe^{26}$ recombination time 
is $t_{rec} \sim 10^{-6}T_7^{1/2}n_{17}^{-1}$ s.  For a mass fraction 
of Fe advected from the core $x_{Fe}$, the number of Fe nuclei in the 
outer Thomson depth unity layer of the bubble (of average depth 
$\tau_T\sim 10^2$) is $N_{Fe}\sim x_{Fe} M_b/(\tau_T 56 m_p) \sim 
4 \times 10^{49}x_{Fe}$.  The Fe line luminosity is 
$(N_{Fe}.\hbox{7 keV} /t_{rec})(1+z)^{-1}$, or  
\beq
L_{Fe}\sim 10^{47} M_{b28}x_{Fe}(2/[1+z])~\ergsi~,
\label{eq:Lfe}
\enq
where $10^{28} M_{b28}$ g is the bubble mass in the absence of entrainment. 
A modestly supersolar Fe mass fraction $x_{Fe}\sim 10^{-2}$ (or less, if 
$M_{b28} >1$ due to entrainment) yields a recombination Fe line luminosity 
comparable to the typical observed value $L_{Fe}\sim 10^{45}\ergsi$ (e.g. 
Piro, \etal., 2001), the mass of Fe involved being $\siml 10^{-5}\msun$. 
More detailed radiative transfer calculations are under way 
(Kallman, \Mesz \& Rees, 2001).

\section{Implications for Progenitors and Fe Lines}
\label{sec:disc}

The He core dimensions do not vary greatly among the different types of 
massive evolved stars, whether they have lost their envelopes or not. 
The exact values depend on uncertain details of wind mass loss, rotation 
rate, binary mass exchange, common envelope evolution, etc. (e.g. Heger, 
\etal, 2001).
Similarly, the density and pressure profiles are roughly $\rho \propto r^{-3}$ 
and $p\propto r^{-4}$, with the density suddenly dropping at the outer edge 
of the He core to $1-10^2\gcmcui$.
Provided that the pressure and density near the center of the precollapse 
star vary more slowly than linearly with the initial total stellar mass, 
the initial properties of the jet injected near the BH horizon would be 
approximately similar to those discussed in \S \ref{sec:jet}. 

Thus, if it is a general property that the jet becomes relativistic near the 
outer radius $r_{He}\sim 10^{11}$ cm of the He core, even for a substantially
extended envelope outside of this, its lab-frame crossing time $r_H/c\Gamma_h^2$ 
adds little to the previously incurred He-core crossing time $t_{He} \sim 10-20$ s. 
Thus, provided the central engine (jet) feeding time exceeds $t_{He}$, 
the jet would be expected to break free of the envelope, and in principle 
lead to a successful GRB.  This could increase the range of potential GRB 
progenitors, including (besides the He-star merger and Wolf-Rayet candidates, 
e.g. MacFadyen \etal, 2000) also blue supergiants, unless the advancing jet 
(depending also on $\theta_f$) sweeps up so much envelope mass that 
$\Gamma_h$ drops drastically.  

Unsuccessful GRB, where the jet is choked before emerging, would occur for 
central engine lifetimes $t_j \siml t_{He}\sim 10~(r_{He}/10^{11}\hbox{cm})
(V_h/c)^{-1}~\hbox{s}$. The TeV neutrino signatures of bursts discussed by 
\Mesz \& Waxman (2001) would thus have a characteristic duration $t_{He}
\sim 10$ s, which in successful GRB would precede the $\gamma$-rays by 
about $t_{He}$ s, while in $\gamma$-ray dark (failed) GRB they might be
the only immediate observable manifestation of a choked collapsar jet.
However, the cocoon of relativistic matter created during the sub-relativistic
jet passage through the star would, in the choked case, not have a ready-made 
escape route, yet still have more energy than the envelope binding energy;  
it could expand more or less isotropically through the envelope, and (as also 
envisaged by Woosley \etal, 2000 and Aloy \etal 2000) violently disrupt it.
For jets which start out with $V_{h,0}\ll 10^8$ cm/s 
the He core crossing time could substantially exceed 10 s, and the total energy 
in the cocoon might be $\simg 10^{52}$ ergs, giving rise to a ``hypernova" 
(\Pacz, 1998, Iwamoto, \etal, 1998, Wang \& Wheeler, 1998), without, however, 
an accompanying GRB. This would appear, after the disrupted envelope becomes 
optically thin, as a peculiar SN Ib/c, of greater than usual brightness.

In a successful jet break-through, not only would a conventional ``long" 
($\simg 10$ s) GRB be detectable, followed by a ``standard" non-thermal 
power-law decay afterglow from the decelerating jet blast wave, but also there 
would be, after hours to a day, a secondary brightening in the light curve, 
caused by the emergence of a bubble with $\sim 10^{51}$ erg. 
A reliable estimate of the amount of Fe entrained in the bubble would require
numerical calculations. However, under plausible conditions, the bubble can 
produce a Fe K-$\alpha$ X-ray line with $l_{Fe}\sim 10^{45}\ergsi$.

\acknowledgements{We thank NASA NAG5-9192, NAG5-9153, the Sackler Foundation, 
the Royal Society, and the referee for comments.}


\begin{references}

\refe{aloy+00} Aloy,M, M\"uller,E, Iba\~nez,J, Mart\'i,J \& MacFadyen,A 2000,ApJ531,L119
\refe{begelman84} Begelman, M, Blandford, R \& Rees, M.J., 1984, Rev.Mod.Phys.56, 255
\refe{funnel76} Fishbone, L \& Moncrief, V, 1976, ApJ 207, 962
\refe{heger+01} Heger, A, MacFadyen, A \& Woosley, S, 2001, in {\it Black Holes in Binaries
 and Galactic Nuclei}, Eds. L. Kaper, \etal (Springer:New York), p.332
\refe{iwa98} Iwamoto, K, \etal, 1998, Nature 395, 672
\refe{kall01} Kallman, T, \Mesz, P \& Rees, M.J., 2001, in preparation
\refe{macfadyenwooheg01} MacFadyen, A, Woosley, S \& Heger, A, 2001, ApJ 550, 372
\refe{meszwax01} \Mesz, P \& Waxman, E, 2001 (astro-ph/0103275)
\refe{narapiku01} Narayan, R, Piran, T \& Kumar, P, 2001, ApJ subm (astro-ph/0103360)
\refe{pac98} \Pacz, B., 1998, ApJ, 494, L45
\refe{piro01} Piro, L, \etal, 2001, review at the April APS Meeting, Washington DC.
\refe{reesmesz00} Rees, M.J. \& \Mesz, P, 2000, ApJ, 545, L73
\refe{scheu74} Scheuer, P, 1974, MNRAS, 166, 513
\refe{wangwhee98} Wang, L \& Wheeler, J.C., 1998, ApJL, 504, L87
\refe{wheeler00} Wheeler, J.C., Yi, I; Hoefflich, P. \& Wang, L., 2000, ApJ 537, 810
\refe{woo+93} Woosley, S, Langer, N \& Weaver, T, 1993, ApJ, 411, 823
\refe{woosley99} Woosley, S., 1999, in {\it Gamma-Ray Bursts}, 
  eds. R. Kippen, \etal, AIP Conf.Proc. 526, p. 555 



\end{references}
\end{document}